\documentclass[pre,twocolumn,showpacs,floatfix,nofootinbib,groupaddress,showkeys]{revtex4-1}

\usepackage{amsmath}
\usepackage{amsfonts}
\usepackage{amssymb} 
\usepackage[margin=0.75in]{geometry}
\usepackage{graphicx} 
\usepackage{cleveref}
\usepackage{bm}
\usepackage[outdir=./]{epstopdf}
\renewcommand{\[}{\begin{equation}}
\renewcommand{\]}{\end{equation}}

\begin{document}

\bibliographystyle{apsrev}
\title{Nucleation of antagonistic organisms and cellular competitions on curved, inflating substrates}
\author{Maxim O. Lavrentovich}
\email{lavrentm@gmail.com}
\affiliation{Department of Physics \& Astronomy, University of Tennessee, Knoxville, Tennessee 37996, USA}
\author{David R. Nelson}
\email{drnelson@fas.harvard.edu}
\affiliation{Department of Physics, Harvard University, Cambridge, Massachusetts 02138, USA}

\begin{abstract}
 We consider the dynamics of spatially-distributed, diffusing populations of organisms with antagonistic interactions. These interactions are found on many length scales, ranging from kilometer-scale  animal range dynamics with selection against hybrids to micron-scale interactions between poison-secreting microbial populations. We find that the dynamical line tension at the interface between antagonistic organisms suppresses survival probabilities of small clonal clusters: the  line tension introduces a critical cluster size that an organism with a selective advantage must achieve before deterministically spreading through the population.
We calculate the survival probability as a function of selective advantage $\delta$ and antagonistic interaction strength $\sigma$. Unlike a simple Darwinian selective advantage, the survival probability depends strongly on the spatial diffusion constant $D_s$ of the strains when $\sigma>0$, with suppressed survival when both species are more motile. Finally, we study the survival probability of a single mutant cell at the frontier of a growing spherical cluster of cells, such as the surface of an avascular spherical tumor. Both the inflation and curvature of the frontier significantly enhance the survival probability by changing the critical size of the nucleating cell cluster.   \end{abstract}
\maketitle

\section{Introduction}

  Antagonistic interactions between species, genotypes, or different organism types within a population are ubiquitous across the tree of life and span many length  and time scales. For example, animal species such as the fire-bellied and yellow-bellied toad can interbreed and create selectively disadvantageous hybrids. The selection against hybrids can stabilize the ranges of these toads, creating narrow interfaces or ``clines'' between the species \cite{toads}.  On much smaller length scales, many Gram-negative bacteria have a type-VI secretion system which allows them to inject nearby cells with ``effector'' molecules that can poison incompatible strains  \cite{type6rev}.  Genetically encoded toxin-antitoxin cassettes confer protection on the poisoners. Like the selection against hybrids, this antagonistic interaction stabilizes linear interfaces between spatial domains or circular boundaries surrounding genetically homogeneous patches of organisms \cite{yunker1}. For a review of social interactions between microorganisms in two dimensions, see Ref.~\cite{ratcliffrev}.

Distinct clones within cell tissues may also have  antagonistic interactions. For example, cancers have a particularly rich ecology \cite{tumor1} due to their heterogeneous genetic composition \cite{bigbang}. Competitive interactions such as \textit{amensalism}, in which a strain can inhibit the growth of nearby strains while remaining unaffected, is likely present amongst the heterogeneous cells of a tumor. Such an interaction may lead to the stabilization of distinct clones in solid tumors \cite{tumor2}. Understanding such ecological effects within the tumor is crucial for developing better cancer treatment options \cite{korolev1}. Similar antagonistic effects are relevant for species with artificial gene constructs, such as the CRISPR-Cas9 technology introduced to contain certain pest populations \cite{pest}. In this case, the effective antagonistic interaction between the engineered  and  wild-type pest strains provides a mechanism for control of the wild-type population  \cite{crispr}. We  model two-dimensional populations of two antagonistic strains, relevant for all of these scenarios. In this paper, we  study not only flat, two-dimensional populations (as one might find in the animal range scenario), but also thin, effectively two-dimensional frontiers of   growing spherical clusters of cells. The latter evolutionary dynamics has implications for antagonism near the frontiers of avascular tumors.

\begin{figure}[htp]
\centering
\includegraphics[width=0.45\textwidth]{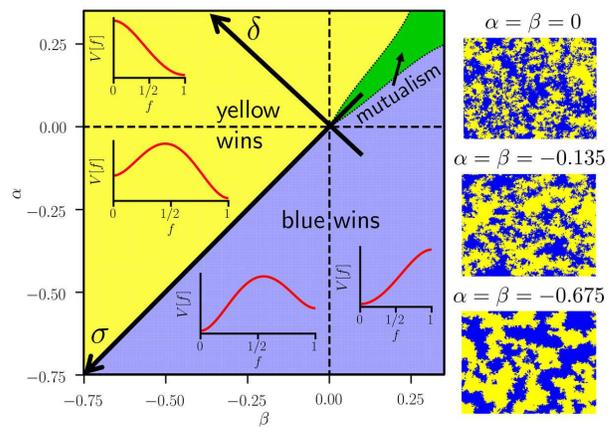}
\caption{\label{fig:intro} Phase diagram sketch for frequency-dependent interactions $\alpha,\beta$  for two-dimensional populations or the dynamics at the flat, leading edge of a three-dimensional range expansion. To explore the third quadrant $(\alpha,\beta<0)$ of this phase diagram, it is convenient to work in rotated coordinates $\delta \equiv \alpha - \beta$ and $\sigma \equiv -(\alpha+\beta)/2$.  When $\alpha,\beta<0$, the line tension  between domains increases with increasing $\sigma>0$. In the  first quadrant ($\alpha,\beta>0$) beneficial interactions lead to a mutualistic phase (green wedge), seen in experiments on range expansions of yeast strains \cite{Mueller}.   The effective potential $V[f]$ for the fraction of yellow strains $f$ is indicated in the various $(\alpha,\beta)$ regimes.  In the third quadrant, this potential has two local minima at $f=0,1$. The dashed lines bounding the third quadrant are limits of metastability for  $V[f]$. On the right we show simulations of an initially well-mixed population of yellow and blue cells in equal proportion, evolved for $t=2000$ generations for three values of $\alpha=\beta \leq 0$. If $\alpha \neq \beta$, one of the strains has a selective advantage and can out-compete the other strain. } 
\end{figure}

 We focus on the dynamics of two species, labelled blue and yellow, with growth rates $\Gamma_{b}$ and $\Gamma_y$, respectively. Consider a monolayer or thin layer of individuals at carrying capacity, such that the fraction of yellow cells is $f$ and the fraction of blue is $1-f$. To model the antagonism, we suppose that the growth rates depend on the local fraction   of the other species. Hence, 
\begin{equation}
\begin{cases}
\Gamma_y = \Gamma_y^0+\alpha(1-f) \\
\Gamma_b = \Gamma_b^0+\beta f
\end{cases}, \label{eq:growthrates}
\end{equation}
where $\Gamma_{y,b}^0$ are base growth rates and the parameters $\alpha$ and $\beta$ are the amplitudes of the fraction-dependent contributions for the yellow and blue strain, respectively \cite{KorolevMut}. We will take  $\alpha,\beta\leq 0$, so that the two strains have mutually antagonistic interactions and suffer a growth penalty when either type grows next to the other. As discussed below, these interactions create a dynamical ``line tension'' $\gamma$ between cell types, as shown in Fig.~\ref{fig:intro}, where yellow and blue strains demix more readily as we make $\alpha,\beta$ more negative. Here we use new coordinates $\delta = \alpha-\beta$ and $\sigma = -(\alpha+\beta)/2$. The line tension $\gamma$ is a monotonically increasing function of $\sigma$, $\gamma \equiv \gamma(\sigma)$, in the third quadrant  of the phase diagram in Fig.~\ref{fig:intro} \cite{MOLMut}.

  The line tension $\gamma$ is generated  by the spatial evolutionary dynamics between strains and describes a tendency of interfaces between domains of yellow and blue strains to minimize their lengths. As we shall see, we can interpret this tendency as an ``energetic penalty'' for creating a larger interface,  analogous to a surface tension term in a free energy describing phase ordering. The presence of the dynamical line tension will allow us to make a connection between the evolutionary dynamics and phase-ordering dynamics of nucleation and growth.  These dynamics become especially interesting on the curved, inflating frontiers of growing tissues such as microspheroidal tumors because the underlying space on which the nucleation and growth dynamics plays out is now curved and time-dependent! Also, note that unlike typical surface tensions, the dynamical line tension $\gamma$ does not depend on mechanical forces such as those driving microbial colony shapes \cite{yeastfluid,hallatschektension}.

 We  also discuss briefly the $\alpha,\beta > 0$ case, i.e., the first quadrant of Fig.~\ref{fig:intro}. Here, we have mutualism where each cell type benefits from the presence of the other type. In this case, there is a possibility of a ``mutualistic phase'' in which the two species remain mixed as the population evolves (shown schematically as a wedge centered on the line $\alpha=\beta$ in the first quadrant of the phase diagram in Fig.~\ref{fig:intro}). In this case, the mutualistic phase has to combat number fluctuations embodied in genetic drift, which can drive the population locally to fixation of one or the other strain. Although mutualism prevails in the entire first quadrant within mean field theory, the region of stable mutualism is reduced to a narrow wedge by these fluctuations. This mutualistic regime has been observed in yeast range expansions \cite{Mueller} and has been the subject of previous theoretical work \cite{KorolevMut,MOLMut}.

We will focus on thin populations and monolayers of individuals in particular. This scenario could describe a thin layer of microorganisms on a Petri dish, a geographical range of animals, or the frontier of a growing  cancerous cell mass.  In the latter case, we assume the growth to be  avascular, with the tumor receiving nutrients only via diffusion from the surrounding medium. Then, only the cells close to the curved frontier will divide and this interface becomes in effect a locally two-dimensional population. Such a growth modality has been observed in microspheroidal tumor cultures \cite{spheroids2}.

 Consider  the local  fraction $f \equiv f(\mathbf{x},t)$ of  yellow cells at some position $\mathbf{x}=(x,y)$  in the population at time $t$. To derive a spatial dynamics for our strains, we employ a ``stepping stone model''  \cite{KimuraWeiss} where we assume that at each position $\mathbf{x}$ the population can be modelled as well-mixed with some characteristic genetic drift strength $D_g \approx a^2/(\tau_g N_{\mathrm{eff}})$, where $N_{\mathrm{eff}}$ is a local population size and $a$ is a characteristic length over which we can treat the population as well-mixed. In our case, we will be concerned primarily with monolayers of cells for which $a$ is a cell diameter and $N_{\mathrm{eff}} \approx 1$. All of the dynamics associated with antagonism and selection embodied by the growth rates in Eq.~\eqref{eq:growthrates} will occur within these well-mixed populations. Then, we assume that cells exhibit some small diffusive motion which leads to a cell exchange between adjacent positions. This means that the fraction $f$ will have a spatial diffusion with a characteristic spatial diffusion constant $D_s$.

Combining the well-mixed dynamics with the spatial diffusion leads to a simple time evolution for the spatial fraction when $\delta$ and $\sigma$ are small compared to the average base growth rate $\langle \Gamma^0 \rangle \equiv (\Gamma_y^0+\Gamma_b^0)/2$ :
\begin{equation}
\partial_t f = D_s \nabla^2 f+  \frac{f(1-f)}{\langle{\Gamma}^0\rangle} \left[\frac{\delta}{2} +\sigma \left(2f-1 \right)\right]+\bar{\eta}, \label{eq:steppingstone}
\end{equation}
where $\sigma=-(\alpha+\beta)/2$ is an antagonistic interaction strength ($\alpha,\beta<0$), $\delta  =  \Gamma_y^0- \Gamma_b^0+\alpha-\beta$ is a selective advantage of the yellow cells,   and $\bar{\eta} \equiv \bar{\eta}(\mathbf{x},t)$ is a stochastic noise due to individual cell births and deaths in each local well-mixed population. 
Since only $\delta$ is influenced by the difference in base growth rates, we will set, without loss of generality,  $\Gamma_b^0=\Gamma_y^0=\tau_g^{-1}$ with $\tau_g$ a generation time. The nonlinear noise is interpreted in the It\^o sense, has a zero average $\langle\bar{\eta} \rangle=0$, and has correlations bilinear in the organism fractions $f(\mathbf{x},t)$ and $1-f(\mathbf{x},t)$,
\begin{equation}
\langle\bar{\eta}(\mathbf{x},t)\bar{\eta}(\mathbf{x}',t') \rangle= D_g f(1-f)\, \delta(\mathbf{x}-\mathbf{x}')\delta(t-t'), \label{eq:noise}
\end{equation}
where $D_g$ is the genetic drift strength mentioned previously.  Details of such stochastic ``stepping stone models'' \cite{KimuraWeiss}, with a focus on one spatial dimension, are reviewed in Ref.~\cite{KorolevRMP}. We shall also see that the dynamics embodied in Eqs.~\eqref{eq:steppingstone} and \eqref{eq:noise} can be interpreted as a noisy evolution in an effective potential $V[f]$, which is sketched in Fig.~\ref{fig:intro} for various $\alpha$ and $\beta$ parameter regimes of interest.
This potential has two local minima at $f=0$ and $f=1$. The dashed lines bounding the third quadrant in Fig.~\ref{fig:intro} are limits of metastability for $V[f]$. 
 
\begin{figure}[htp]
\centering
\includegraphics[width=0.4\textwidth]{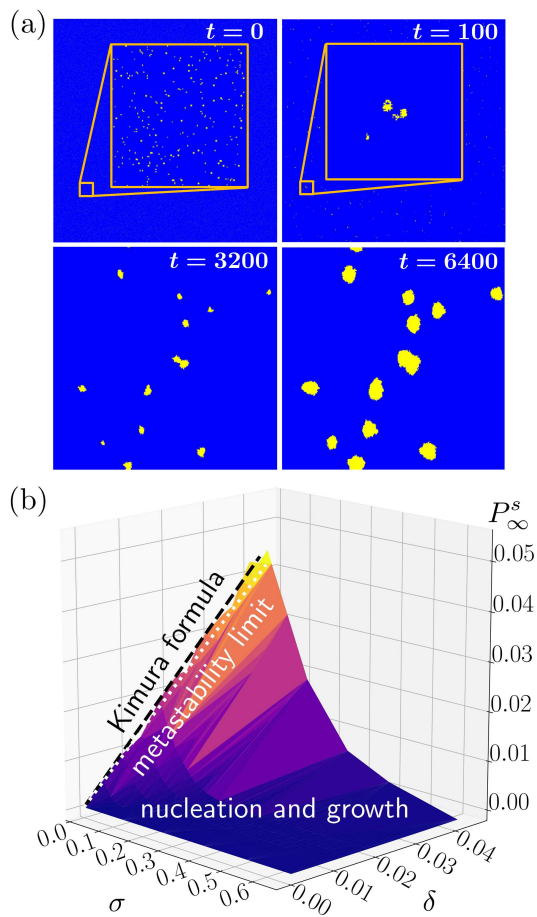}
\caption{\label{fig:nucleation} (a) Evolution of  a dilute suspension of yellow cells (at an initial fraction $f_0=0.02$) in units of the generation time $\tau_g$, in a population of blue cells for antagonism strength $\sigma=0.675$ and a small selective advantage $\delta=0.045$, given in units of $\tau_g^{-1}$. The total simulated population is a monolayer of 2048 by 2048 cells with periodic boundary conditions. With these parameters, the yellow cells, even though they have a selective advantage, must create a roughly circular cluster larger than a critical size of about $40$ cells in order to be able to generate a spreading population. Note the very similar pattern of yellow clusters in the last two frames, showing that essentially all the yellow clusters exceed the critical size when $t=3200$.  (b) Simulated survival probability $P_{\infty}^s$ of a single yellow cell in a sea of blue at long times for the indicated values of $\sigma$ and $\delta$. The surface is generated by tessellating the simulation points $(\sigma,\delta,P_{\infty}^s)$. The black dashed line shows the Kimura formula in Eq.~\eqref{eq:kimura}, evaluated in the limit of a large system size. The white dashed line indicates the boundary ($\sigma=\delta/2$) at which the yellow strain state becomes metastable for smaller values of $\sigma$ or larger values of $\delta$ (i.e., this is the top boundary of the third quadrant in the phase diagram in  Fig.~\ref{fig:intro}).   }
\end{figure}

 As is typical of non-equilibrium systems, the initial condition  for the populations plays a crucial role. Unlike equilibrium systems which evolve to a unique  state fully specified by a free energy,  the nature of the initial distribution of strains within the population can  dramatically influence the subsequent evolutionary dynamics. In this paper we will focus on two kinds of initial conditions: (1) a random mixture of yellow and blue cells with equal concentrations [$\langle f(\mathbf{x},t=0) \rangle_{\mathbf{x}}=1/2$, where $\langle \cdot \rangle_{\mathbf{x}}$ represents a spatial average]  which coarsens over time as illustrated in Fig.~\ref{fig:intro}, creating genetic ``sectors'' whose typical size grows in time, and (2) a  dilute suspension of yellow cells with a selective advantage $\delta >0$ in an otherwise all-blue population, as shown in Fig.~\ref{fig:nucleation}(a). The first initial condition is analogous to spinodal decomposition \cite{bray}, and would arise from an initial condition of genetic mixing in the mutualistic wedge in the first quadrant of Fig.~\ref{fig:intro} with $\alpha \approx \beta >0$, if the population were suddenly moved to an environment with $\alpha \approx \beta < 0$.

In the first scenario, strains can  compete with  each other locally and the mixture coarsens over time. This behavior is typical in microbial studies where such an initial condition may be generated by inoculating, say, a well-mixed planar array of bacteria without flagella  on a Petri dish \cite{yunker1}.  We will first study the coarsening properties of the domains in this  scenario. In this case, we assume cells that are killed off by antagonism give up their volume by bursting, with contents that are either absorbed into the  agar gel, or provide nutrients to the antagonists. The phase diagram describing the long-time dynamics as a function of $\sigma$ and $\delta$ for a random mixture of blue and yellow cells in equal proportions is sketched in Fig.~\ref{fig:intro}. When $\delta>0$, the yellow cells will eventually sweep the population. When $\delta < 0$, the blue cells are victorious. We will study the coarsening dynamics near the line $\delta=0$, and study the effect of increasing $\sigma$.

In the second scenario, yellow cells (representing, e.g., rare mutants within the population), if they survive genetic drift, can out-compete neighboring blue cells when their selective advantage $\delta>0$. If the resultant yellow cluster  reaches a certain critical cluster size, it can spread through the blue population, as shown in Fig.~\ref{fig:nucleation}(a). We will connect such a dynamics to a nucleation and growth process where the blue population represents a metastable state and the yellow cells correspond to a stable, nucleating minority phase. This connection is possible because Eq.~\eqref{eq:steppingstone}, apart from the noise correlations in Eq.~\eqref{eq:noise}, has an identical form to the stochastic partial differential equation describing non-conserved, model A dynamics at a first-order phase transition \cite{bray,hohenberghalperin}.  We shall now briefly comment on the difference in the noise.

In the typical (model A) nucleation and growth scenario, such as in a kinetic Ising model in two dimensions, or  near a two-dimensional liquid-vapor phase transition, the noise comes from thermal fluctuations throughout the system and has a simple correlation: The nonlinearity  $f(1-f)$ in Eq.~\eqref{eq:noise} is replaced by a  constant proportional to the temperature.  In contrast, for our antagonistic evolutionary dynamics, the noise has nonlinear correlations and acts \textit{only close to interfaces} between ``phases'' of $f=1$ (yellow cells) and $f=0$ (blue cells).  
However,  when interfaces between the two phases are \textit{sharp} (the so-called ``thin-wall limit''), thermal noise in model A primarily acts to distort the \textit{position} of the interface: the low-energy  modes excited by thermal noise are translations or undulations of the nucleating droplet boundary \cite{voloshin,Gunther}, as shown schematically in Fig.~\ref{fig:droplet}(b).  In one spatial dimension, it can be shown rigorously that the interface dynamics behaves like a random walk biased by the asymmetry in potential well depth \cite{mathinterface}. Hence, it seems reasonable to conjecture that the model A scenario and the antagonistic dynamics  are analogous in the thin-wall limit for single ``droplets'' of the  stable phase (yellow cells) surrounded by the metastable phase (blue cells). Specifically, the dynamics described by Eq.~\eqref{eq:steppingstone} should reduce, in this thin-wall case, to the dynamics of  a fluctuating, sharp  interface. Thus, for the problems of interest to us here, it seems plausible that we can evaluate the  nonlinear noise correlations in Eq.~\eqref{eq:noise} at the middle of the interface where $f=1/2$, a conjecture we check with our numerical simulations below.

It is also worth noting that in the case $\sigma=0$ and variable selective advantage $\delta$, the survival probability $P^s_{\infty}$ at long times $(t \rightarrow \infty)$ of a single yellow cell in a sea of blue is given by the celebrated Kimura formula, which, although originally derived for well-mixed populations, is remarkably independent of the spatial structure of the population under rather general conditions \cite{maruyama}. For a single cell in a population of $N \gg 1$ blue cells, this formula reads:
\begin{align}
P^s_{\infty} &  = \frac{1- \exp\left[- \tau_g\delta/2\right]}{1- \exp\left[{-   \tau_gN\delta/2}\right]} \nonumber \\ & \approx 1-\exp\left[ -\frac{\delta }{D_g}\int\mathrm{d}^2\mathbf{x}\,f(\mathbf{x},t=0)\right] \approx\frac{A_0 \delta  }{D_g}, \label{eq:kimura}
\end{align}
where $\tau_g$ is the generation time. The first line is the original formula of Kimura \cite{kimurabook}, and the second line is its generalization to two dimensions with $\sigma=0$ (no antagonism). The first approximate equality only holds for $N$ so large that $ \tau_g N \delta \gg 1$. Note the remarkable independence of this result on the precise spatial arrangement of yellow cells at $t=0$, $f(\mathbf{x},0)$. The initial area occupied in the case of an isolated yellow cluster is $\int \mathrm{d}^2\mathbf{x} \,f(\mathbf{x},t=0)=A_0$ ($A_0 \approx a^2$ for a single cell with diameter $a$). The last approximate equality in Eq.~\eqref{eq:kimura} requires a small selective advantage, $\delta \ll D_g/A_0$,  which will be our regime of interest. Because $D_g \sim a^2/(\tau_g N_{\mathrm{eff}})$,  we require $\delta \ll \tau_g^{-1}$ for the special case of a single mutant cell and $N_{\mathrm{eff}}=1$.  Note also that this formula is \textit{independent of the spatial diffusion coefficient} $D_s$. This  is a general phenomenon and has been checked for a wide range of systems  \cite{maruyama,doering,pigolotti}.

 We plot Eq.~\eqref{eq:kimura}  on one of the panels of Fig.~\ref{fig:nucleation}(b), where it clearly fits our simulation data (described below) for $\sigma=0$ (with $\tau_g D_g/A_0=1.00\pm0.01$ as a fitting parameter). We shall see that when we include an antagonistic interaction, $\sigma>0$, the survival probability of a single mutant cell (or small patch of cells) as a function of selective advantage $\delta$ is strongly suppressed, as illustrated in Fig.~\ref{fig:nucleation}(b). Moreover, the formula for $P^s_{\infty}$ is qualitatively different for $\sigma>0$ and depends explicitly on the spatial diffusion constant $D_s$ and the spatial structure of the population.

The paper is organized as follows: In Section~\ref{sec:setup}, we describe our simulations and analytic approaches. In Section~\ref{sec:coarsen}, we consider how domains of yellow and blue cells coarsen over time for $\delta=0$ and $\sigma \geq 0$, i.e., right along the line of first-order transitions in the third quadrant of Fig. 1. We show that increasing $\sigma$ increases the line tension between yellow and blue cell clusters, as illustrated in the population snapshots in Fig.~\ref{fig:intro}. In the Section~\ref{sec:survival}, we analyze the probability that a single yellow cell surrounded by blue cells is able to create a  cluster that survives and eventually expands at long times.  In Section~\ref{sec:inflation}, we extend our analysis of the survival probability to range expansions with inflating spherical frontiers, such as the surface of an avascular solid tumor. We conclude with some discussion of our results and prospects for future work in Section~\ref{sec:conclusion}.

\section{Simulations and Analysis \label{sec:setup}}

\begin{figure}[htp]
\centering
\includegraphics[width=0.48\textwidth]{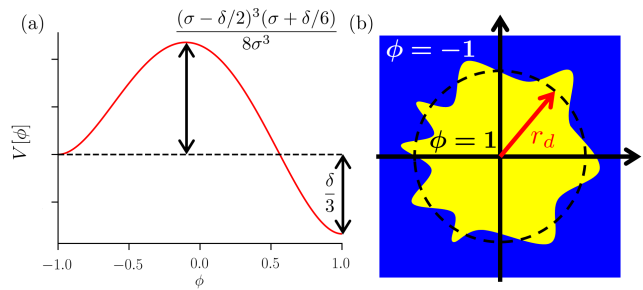}
\caption{\label{fig:droplet} (a) The stochastic equation [Eq.~\eqref{eq:diffpot}] governing the evolutionary dynamics can  be interpreted as a (spatial) escape over a barrier problem with the indicated potential $V[\phi]$. Note that the barrier height is proportional to $\sigma$ for $\sigma \gg \delta$, while $\delta $ sets the difference between the two minima. (b) The nucleation over the barrier occurs when a disc-shaped cluster forms with a radius $r_d$ larger than a certain critical size $r_d>r_d^*$. By studying fluctuations around the critical cluster (dashed line), we are able to calculate the escape rate $\Gamma$ at which clumps of yellow cells close to this critical size are able to nucleate and grow. As discussed in detail in Section~\ref{sec:survival},  we work in the thin-wall limit $r_d \gg \sqrt{D_s/\sigma}$, where the boundary between the yellow and blue cells is \textit{sharp} and the nonlinear noise in Eq.~\eqref{eq:noise} produces undulations in the  interface position as shown. }
\end{figure}

To begin, it is convenient to make some connections to models describing the relaxation of an (Ising-like) magnetic spin system to equilibrium. To do this, we change variables from $f$ to $\phi=2f-1$, so that the yellow and blue strains correspond to $\phi = \pm 1$, respectively.  With this variable change, the equation of motion in Eq.~\eqref{eq:steppingstone} is analogous to  the dynamics of an Ising-like system well below the critical temperature $T_c$, with a coarse-grained ``magnetism'' $\phi$ evolving in a potential $V[\phi]$ which exhibits two minima at   $\phi = \pm 1$   (e.g., the two magnetizations of an Ising spin) in such a way that the total magnetization of the system is \textit{non-conserved} (model A) \cite{hohenberghalperin}.   Note that, although competing cell populations are clearly a form of ``active matter'' \cite{marchettiRMP}, the dynamics of motile bacterial and Janus particles that motivate many of these studies do not involve cell divisions, and hence are better described by variants of model B dynamics, with a conserved order parameter \cite{hohenberghalperin}. In our evolutionary dynamics, model A time evolution  means that the total fraction of yellow and blue cells is not conserved, and instead varies in time. The equation of motion in Eq.~\eqref{eq:steppingstone} is now conveniently rewritten as
\begin{align}
\partial_t \phi & = D_s \left[\nabla^2 \phi- \frac{1}{D_s}\frac{dV}{d\phi}\right]+\eta \label{eq:diffpot}  \\
  \langle \eta (\mathbf{x},t) \eta(\mathbf{x}',t) \rangle &= D_g(1-\phi^2) \delta(\mathbf{x}-\mathbf{x}') \delta(t-t'),  \label{eq:diffnoise}
\end{align}
where $\eta = 2 \bar{\eta}$ is again a nonlinear noise and $V[\phi]=   (\phi +1)^2 \left[2 \delta  (\phi -2)+3 \sigma  (\phi -1)^2\right]/24$ has a double-well structure for $\sigma > \delta/2 > 0$, as shown in Fig.~\ref{fig:droplet}(a) and in Fig.~\ref{fig:intro}.
In this regime, the blue state becomes metastable. The  double-well structure of the potential allows us to naturally connect the antagonistic dynamics to nucleation and growth, a connection that was made previously in the ecological context by Rouhani and Barton \cite{gennucleation1} and in the context of gene drives \cite{crispr}.  Although it is tempting to interpret the limit $\sigma \rightarrow 0$ in Fig.~\ref{fig:droplet} in terms of an Ising model in a magnetic field just below its critical point, this is precisely where the unusual noise correlations of Eq.~\eqref{eq:noise}  become important, and produce the distinctly \textit{non}-Ising model features in the first quadrant of the ``phase diagram'' shown in Fig.~\ref{fig:intro}.

How shall we deal with the peculiar noise correlations $\langle \eta(\mathbf{x},t) \eta(\mathbf{x}',t') \rangle= D_g (1-\phi^2) \, \delta(\mathbf{x}-\mathbf{x}')\delta(t-t')$ in Eq.~\eqref{eq:diffnoise}, which differs from conventional model A dynamics \cite{hohenberghalperin}, even deep in the third quadrant? Although the noise $\eta$ here is manifestly non-thermal and provided by the intrinsic birth-death dynamics of the individual cells, it will nevertheless be helpful for us to think about Eq.~\eqref{eq:diffpot} as the dynamics of a field $\phi$ as it relaxes to a minimum of the following  ``free energy'':\[
E[\phi] \equiv \int\mathrm{d}^2\mathbf{x}\left[ \frac{1}{2}(\nabla \phi)^2+\frac{V[\phi]}{D_s}\right] \label{eq:freeenergy}.
\]
The field $\eta(\mathbf{x},t)$, then, can be thought of as a thermal noise that acts \textit{only at interfaces} between domains of $\phi = \pm 1$.  At these interfaces, we have  $\phi\approx0$ and the noise correlations in Eq.~\eqref{eq:diffnoise} are approximately independent of $\phi$. Here, we can connect the noise to an inverse temperature via a formula analogous to the fluctuation-dissipation theorem:
\[
\beta \equiv \frac{2 D_s}{D_g}, \label{eq:effT}
\] so that near domain interfaces, we have    $\langle \eta(\mathbf{x},t) \eta(\mathbf{x}',t') \rangle \approx 2 \beta^{-1} D_s  \delta(\mathbf{x}-\mathbf{x}')\delta(t-t')$ \cite{hohenberghalperin,tauberbook}.  We shall test this approximation further in Section~\ref{sec:survival}. We may now draw analogies between  relaxation dynamics of a magnetic spin system and the evolutionary dynamics.

\begin{figure}[htp]
\centering
\includegraphics[width=0.48\textwidth]{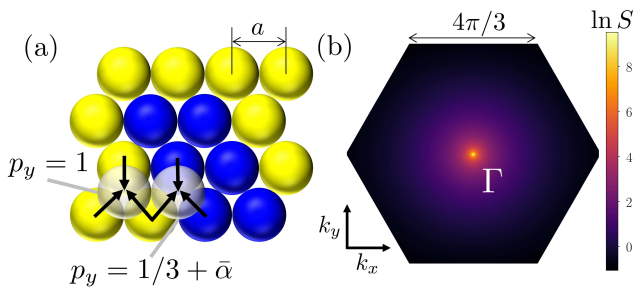}
\caption{\label{fig:sim} (a) A two-dimensional population of  spherical cells with diameters $a$ arrayed as a triangular lattice, evolved one generation at a time by staggering the lattice in each subsequent generation and allowing for three cells to compete to divide into each empty spot. We show two examples of such evolutions, where $p_y$ is the probability that the empty site is yellow. The quantity $\bar{\alpha}$ is related to the selective advantage and antagonism parameters $\delta$ and $\sigma$ of our competition model by $\bar{\alpha}=\delta/9-2\sigma/27$.   (b) The structure factor $S(\mathbf{k},t)$ at $t=16000$ generations for a simulation with an initially well-mixed blue and yellow population with $\sigma=\delta=0$ (voter model). Note that the structure factor close to the origin, which characterizes large distance correlations, is isotropic (depends only on the distance $k \equiv|\mathbf{k}|$ away from the origin $\Gamma$) and has a peak at $k=0$. We choose our units such that the cell diameter $a=1$. For a triangular lattice of organism positions, $S(\mathbf{k},t)$ is defined on a hexagonal Brillouin zone.}
\end{figure}

We  also implement simulations of monolayers of cells, growing either in a two-dimensional triangular lattice or at the edge of an inflating spherical clump of cells. The simulation approaches will be similar to the ones employed in Refs.~\cite{MOLMut,MOLSphere}, which we review briefly here. We evolve  cells one cell division at a time, either in a triangular lattice or on the surface of a spherical cluster of cells generated using the Bennett hard sphere packing algorithm \cite{bennett}.  During each division, three adjacent cells compete to divide into the new spot, as shown in Fig.~\ref{fig:sim}(a). It is convenient to parameterize as follows in order to have the simulations best approximate our choice of parameters in Eqs.~(\ref{eq:growthrates},\ref{eq:steppingstone}): If two cells are blue and one is yellow, then the probability the cell is yellow is $p_y=1/3+\bar{\alpha}$, with $\bar{\alpha}\equiv\delta/9-2\sigma/27$. Otherwise, if we have two yellow and one blue cell, $p_y=2/3-\bar{\beta}$, where $\bar{\beta}\equiv-\delta/9-2\sigma/27$, where $\delta$ and $\sigma$ are in units of $\tau_g^{-1}$. All other possibilities follow from probability conservation. Because we assume no spontaneous mutations on the time scale of our simulations, the daughter cell that appears in a pocket formed by three adjacent cells  of the same color will have the same color. Some examples are illustrated in Fig.~\ref{fig:sim}(a). For the triangular lattice, we keep dividing cells to fill in a staggered copy of the lattice, representing the next generation of cells. For the spherical cluster growth, a generation time $\tau_g$ passes when the total spherical cluster grows by a full cell diameter $a$ .   

An important quantity we  use to characterize the coarsening dynamics in the next section is the structure factor, defined as
\begin{equation}
S(\mathbf{k},t) = \sum_{\mathbf{x}} \langle \phi(\mathbf{y},t) \phi(\mathbf{y}+\mathbf{x},t) \rangle_{\mathbf{y}} \,e^{i \mathbf{k} \cdot \mathbf{x}},
\end{equation}
with $\mathbf{k}=(k_x,k_y)$ a 2-dimensional wavevector and $\langle \ldots \rangle_{\mathbf{y}}$ is a spatial average and an average over many simulation runs. For the triangular lattice with each cell a diameter $a$ away from its six nearest neighbors, $\mathbf{k}$ ranges over a hexagonal Brillouin  zone as shown in Fig.~\ref{fig:sim}(b). If we have an initially isotropic distribution of organism types, we then expect that $S(\mathbf{k},t)$ will also be isotropic during the evolution of the population in time and hence only dependent on $k \equiv |\mathbf{k}|$ at least close to the origin, where we expect details of the lattice structure to drop out. We may use $S(k,t)$ to estimate the characteristic size $\xi(t)$ of strain domains at time $t$.  As $S(k,t)$ is expected to evolve as in a system with non-conserved dynamics, we anticipate that it develops a peak at $k=0$  [see Fig.~\ref{fig:sim}(b)] which will grow and sharpen over time  \cite{bray}. Hence, we can estimate a domain  size $\xi(t)$ from   the half-width at half maximum of the structure factor peak at $k=0$, $k_{\mathrm{HWHM}}$, where $k_{\mathrm{HWHM}} (t) \approx 2\pi/\xi(t)$. 

We can also estimate the average interface density $\rho_{\mathrm{int.}}$ for a population of $N$\ cells using the structure factor. If  the cells are close-packed on some lattice with cell-diameter spacing $a$, then the density of interfaces $\rho_{\mathrm{int}.}$ at time $t$ is given by
\begin{align}
\rho_{\mathrm{int.}} &\propto \frac{1}{ a^2Nz} \sum_{\mathbf{x},\bm{\delta}} \left\langle [\phi(\mathbf{x})-\phi(\mathbf{x}+\bm{\delta})]^2\right\rangle  \nonumber \\ & \propto -\nabla^2S(\mathbf{x}=0,t)\propto \sum_{\mathbf{k}}|\mathbf{k}|^2S(\mathbf{k},t), \label{eq:interfaces}
\end{align}
where  $\langle \ldots \rangle$ is an average over many simulation runs and we sum over all cells at positions $\mathbf{x}$ and their $z$ nearest neighbors at relative positions    $\bm{\delta}$. Note that the summand in the first line of Eq.~\eqref{eq:interfaces} is only non-zero when $\bm{\delta}$ spans an interface between organism types. In the second line, we take the continuum limit    $|\bm{\delta}| \rightarrow 0$ and use a finite-difference approximation to the Laplacian, which can be written in Fourier space as a sum over wave-vectors $\mathbf{k}$ in the  Brillouin zone (BZ) in Fig.~\ref{fig:sim}(b). A convenient quantity, then, measurable in either experiments or simulations in which we uniformly sample $N_{\mathbf{k}}$ different wave-vectors (over the Brillouin zone, say) reads:
\begin{equation}
N_S(t) \equiv\frac{1}{N_{\mathbf{k}}} \sum_{\mathbf{k}}|\mathbf{k}|^2S(\mathbf{k},t) \propto \rho_{\mathrm{int}.}(t). \label{eq:NS}
\end{equation}
The constant of proportionality will depend on the particular lattice or microscopic implementation of the simulated evolution.     We  now proceed to the results of our simulation and analytic approaches.

\section{Domain Coarsening \label{sec:coarsen}}

 \begin{figure}[htp]
\centering
\includegraphics[width=0.4\textwidth]{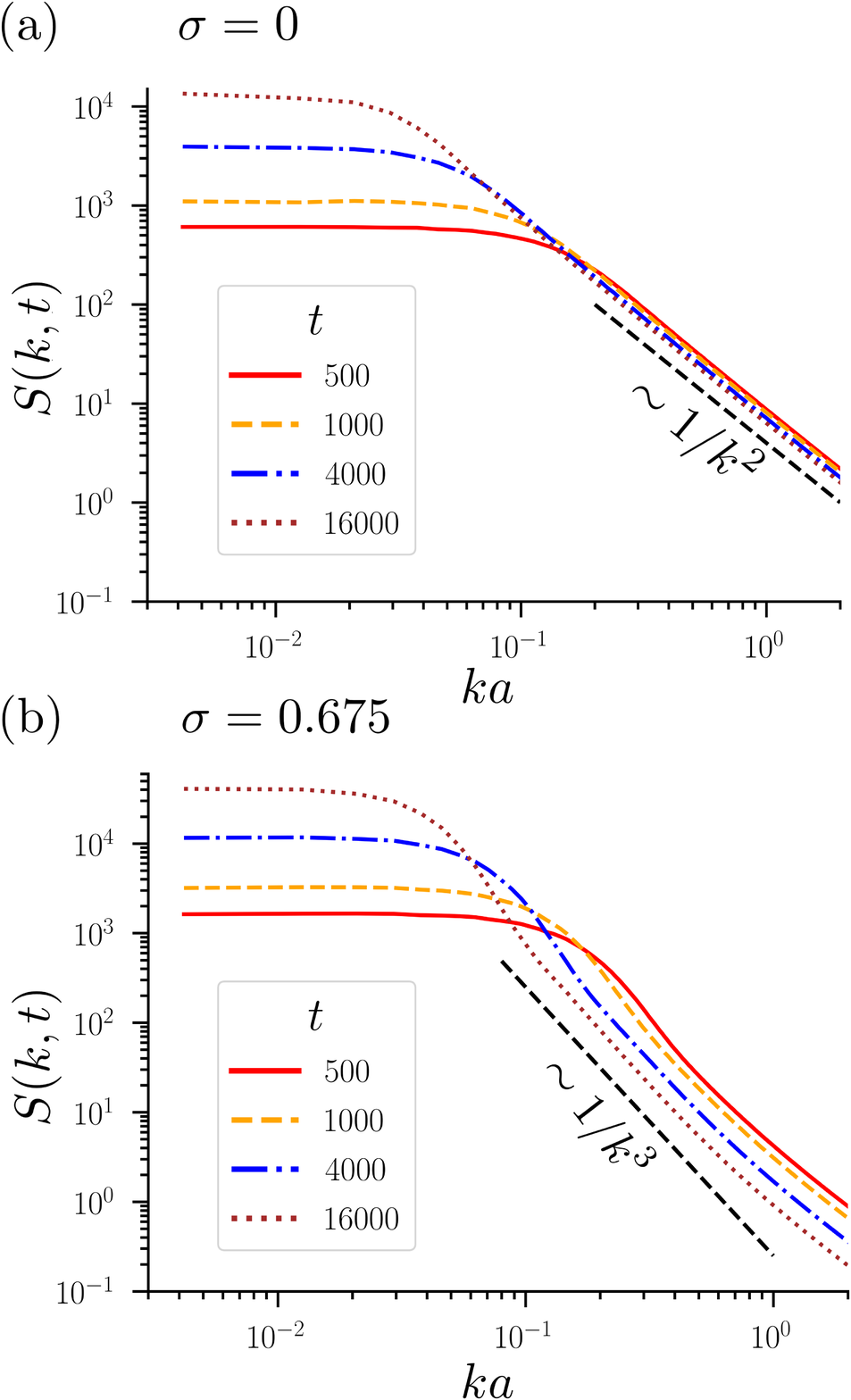}
\caption{\label{fig:SK} Log-log plots of the structure factor $S(k,t)$  $(k=|\mathbf{k}|)$ for different times $t$ (in generations) for an initially well-mixed population of yellow and blue cells in equal proportion with no selective advantage $(\delta =0)$. In (a), $\sigma=0$ and the system exhibits voter-model coarsening dynamics, characterized by diffuse interfaces, which in Fourier space creates the $1/k^2$ tail in the structure factor.\ Conversely, when we have antagonism as in (b), where $\sigma=0.675$, we find Ising-model coarsening with sharp domain walls and a $1/k^3$ tail. Note that in both cases, the structure factor has a peak at $k=0$, which increases and sharpens in time. We use the half-width at half the peak maximum, $k_{\mathrm{HWHM}}$, to estimate the characteristic size of coarsening domains (see Fig.~\ref{fig:coarsening}). } 
\end{figure}

 We first consider a well-mixed initial condition with equal proportions of blue and yellow organisms and ask how the domains of blue and yellow coarsen over time with no selective advantages $(\delta=0)$ but variable antagonism $\sigma$ (as in Fig.~\ref{fig:intro}). The structure factor is indeed isotropic at long wavelengths, as shown in   Fig.~\ref{fig:sim}(b).    When $\sigma=0$, the dynamics of the system reduce to the voter model \cite{MOLMut}, at which we get a peculiar ``line-tension-less'' coarsening. In fact, the full structure factor for the voter model can be calculated exactly and matches our results for $\sigma=0$  \cite{voter1}.  In this case, although the largest domain size in the system grows as $\sqrt{t}$,  the density of interfaces decays with a much different (and slower) $1/\ln t$ behavior. The latter slow decay is due to the presence of very diffuse domains in the system. The  behavior of $S(k,t)$ at larger wavevectors $k$ probes the interface topology. Specifically, for \textit{sharp} interfaces, we expect according to Porod's law that $S(k,t) \sim 1/k^{3}$ for large $k$ (for a two-dimensional system) \cite{bray}. This is indeed what happens for $\sigma>0$ when we have antagonistic interactions, as shown in Fig.~\ref{fig:SK}(b).  In the voter model case $\sigma=0$ in Fig.~\ref{fig:SK}(a), we see that $S(k,t) \sim 1/k^2$ for large $k$, instead, consistent with the absence of sharp interfaces between domains and suggesting that the contribution to $S(k,t)$ for large $k$  is dominated by small, isolated cell clusters, as can be seen in the $\sigma=0$ panel of Fig.~\ref{fig:intro}.

  \begin{figure}[htp]
\centering
\includegraphics[width=0.48\textwidth]{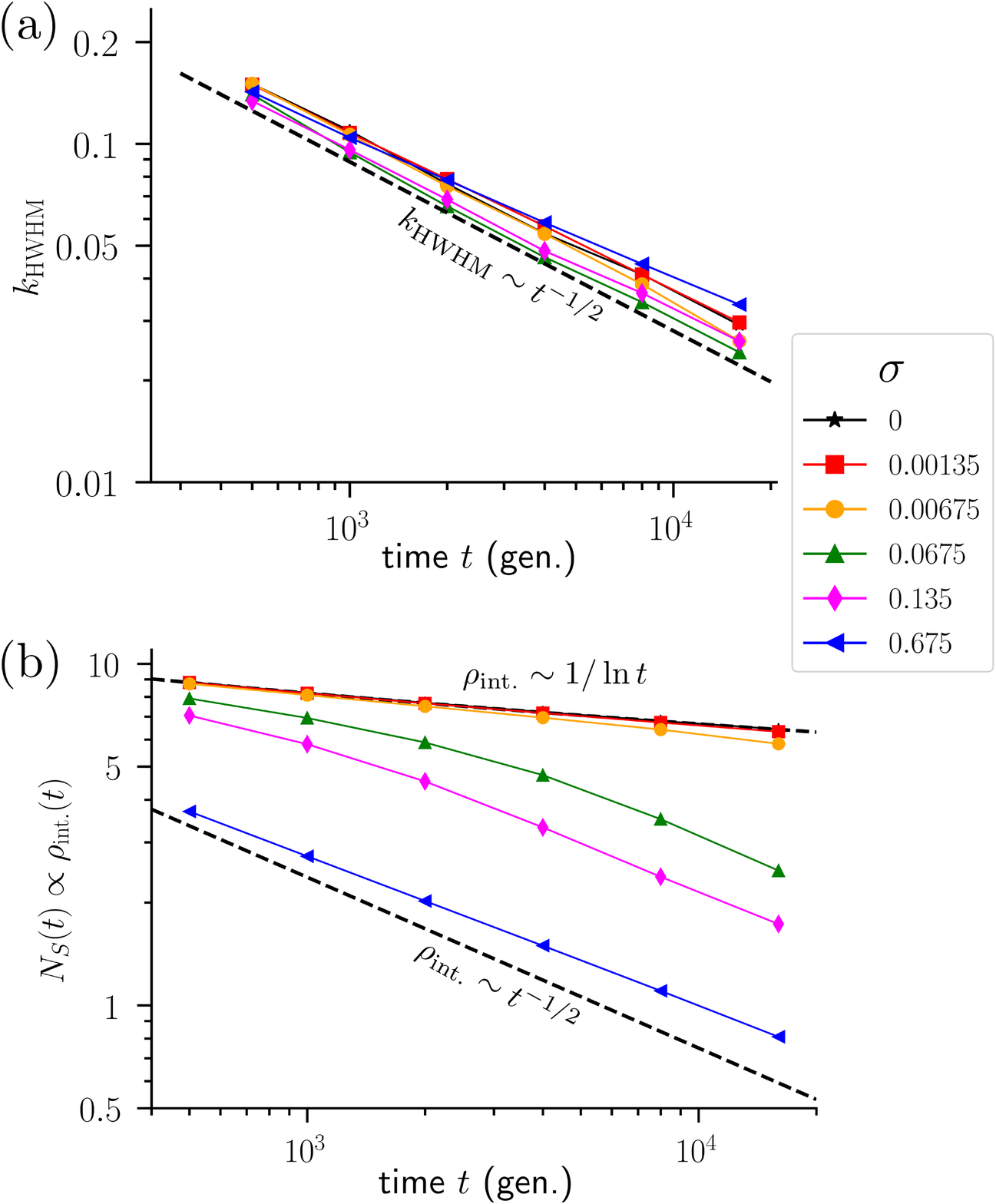}
\caption{\label{fig:coarsening} (a) Half-width of the structure factor peak  (at $k=0$) evaluated at half the maximum (HWHM) as a function of time $t$ (in generations) for various antagonism strengths $\sigma$. Note that for all values of $\sigma$, the width decreases proportionally to $1/\sqrt{t}$, consistent with a typical domain size coarsening as $\xi \sim k_{\mathrm{HWHM}}^{-1}\propto \sqrt{t}$. (b) The quantity  $N_S(t)$, defined in Eq.~\eqref{eq:NS}, is proportional to the interface density $\rho_{\mathrm{int}.}(t)$. As $\sigma$ increases, the density  transitions from a $1/\ln t$ decay to a much faster $1/\sqrt{t}$ decay at longer times, consistent with a sharpening of domain walls with separation $\sim \sqrt{t}$. The lines connect the simulation points.     }
\end{figure}

To detect this  change in the character of the domain interfaces between $\sigma=0$ and $\sigma>0$, we look  at both the interface density $\rho_{\mathrm{int.}}$ and the typical domain size $\xi$ [as judged from $N_S(t)$ in Eq.~\eqref{eq:NS} and the inverse of the half-width at half-maximum $k_{\mathrm{HWHM}}$ of the structure factor peak at $k=0$, respectively] as functions of time $t$. We observe in Fig.~\ref{fig:coarsening} the marked difference between these two quantities. The characteristic domain size $\xi(t)$ always increases as $\xi \propto \sqrt{t}$, with the constant of proportionality only weakly dependent on the antagonistic interaction strength $\sigma$, as indicated by the substantial overlap between the various lines in Fig.~\ref{fig:coarsening}(a). In sharp contrast, the interface density $\rho_{\mathrm{int.}}$ decay depends very strongly on $\sigma$. As $\sigma$ is increased, the line tension at the boundary increases and the interfaces between the domains \textit{sharpen}. This leads to a transition from a $1/\ln t$ decay for $\sigma=0$ to a much faster $1/\sqrt{t}$ decay for $\sigma>0$, as shown in Fig.~\ref{fig:coarsening}(b). This increasing line tension with $\sigma$  also has a pronounced effect on the survival probability of organism genotypes, as we will see in the next section.

\section{Single Cluster Dynamics  \label{sec:survival}}

Let us now consider the fate of a single yellow cell in a (very large) blue cell population, representing some rare mutation, for example. When $\sigma=0$ and the yellow cell has a small selective advantage $\delta>0$, then the probability that it sweeps the population is given by the formula in Eq.~\eqref{eq:kimura}.
However, as soon as $\sigma>0$ and there are antagonistic interactions, the effective line tension  discussed in the previous section strongly suppresses the survival probability, as is evident in Fig.~\ref{fig:nucleation}(b). This suppression can be understood by noting that for $\sigma > \delta/2$, the potential shown in Fig.~\ref{fig:droplet} develops a \textit{barrier} for transitioning between the blue state ($\phi=-1$) and the more favorable yellow type $(\phi=1)$. Similar to conventional nucleation theory, the yellow cell must eventually generate a sufficiently large cluster of offspring to overcome this barrier.

We first consider the deterministic dynamics of a yellow cell cluster   for  $\sigma>\delta/2$. That is, consider a ``droplet'' solution to the dynamical equation Eq.~\eqref{eq:diffpot} in the absence of noise $(\eta = 0)$. If the droplet interface width $\sim \sqrt{D_s/\sigma}$ is much smaller than the droplet radius $r_d(t)$ (see Fig.~\ref{fig:droplet}), then we can look for solutions $\phi(r,t) \equiv \phi^*[r-r_d(t)]$, in which the cluster maintains a stationary  interface profile $\phi^*(z)$ and has a time-dependent radius $r_d(t)$. Under the assumption that the interface profile does not change as the yellow cell cluster expands or shrinks, the profile $\phi^*(z)$ can be estimated from the stationary solution to Eq.~\eqref{eq:diffpot}:
\begin{equation}
 \nabla^2 \phi \approx \, \frac{\partial^2 \phi^*}{\partial z^2} =\frac{1}{D_s}  \frac{d V}{d \phi}[\phi=\phi^*],
\end{equation}
which, under the appropriate boundary conditions [$\phi(z \rightarrow -\infty)=1$ and $\phi(z \rightarrow +\infty)=-1$] is solved by
\begin{equation}
\phi^*(z) =-\tanh \left[ \sqrt{\frac{\sigma }{4D_s}} \,z \right], \quad z=r-r_d(t). \label{eq:profile}
\end{equation}
This solution holds provided the variation of $\phi^*(z)$ is localized around $z=r-r_d=0$ and the cluster radius $r_d$ is large compared to the interfacial width: $r_d \gg \sqrt{D_s/\sigma}$. In this thin-wall limit, we can use Eq.~\eqref{eq:profile} to recast our free energy in Eq.~\eqref{eq:freeenergy} as an energy for a disc-shaped cluster 
\begin{equation}
E[r_d]=2\pi \gamma r_d-\pi c r_d^2 , \label{eq:denergyflat}
\end{equation}
 with a line tension $\gamma=(2/3)\sqrt{ \sigma/D_s}$ and a condensation energy $c=\delta/(3D_s)$.\ Moreover, the dynamics in Eq.~\eqref{eq:diffpot}  becomes \cite{bray} \begin{equation}
 \frac{d r_d}{dt}=- \frac{D_s}{r_d}+\frac{\delta}{2} \sqrt{\frac{D_s}{\sigma}}. \label{eq:dropletd}
\end{equation}
 
There are two important aspects of the Eq.~\eqref{eq:dropletd} dynamics. 
First, when $\delta=0$ (so that the competing organisms are selectively neutral), the droplet nevertheless shrinks and we get the solution $r_d(t)=\sqrt{(r_d^0)^2-2D_s t}$, with $r_d^0$  the initial cluster radius. This solution suggests that curved clusters will have radii that shrink as $\sqrt{D_s t}$ as we would expect from the non-conserved, Ising-like Allen-Cahn (model A) dynamics \cite{allencahn}. Note that the dynamics is independent of $\sigma$, consistent with the results for the domain size $\xi(t)$ we saw in Fig.~\ref{fig:coarsening}(a).
Second, there is  a stationary radius $r_d^*$ at which $dr_d/dt=0$. This is the critical droplet size: If the initial $r_d(t=0)$ is larger than this size, then the two-dimensional droplet grows systematically larger with time, with eventually $r_d(t) \approx (\delta/2) \sqrt{D_s/\sigma}\, t$, where the interface can be regarded as a pushed genetic wave \cite{toads,crispr}. Otherwise, the droplet shrinks. Upon setting $dr_d/dt=0$, we find $r_d^*=\gamma/c=\sqrt{4D_s \sigma/\delta^2}$.
Therefore, in order for a yellow cell cluster to survive,  genetic drift [i.e., the fluctuations embodied in the interfacial  noise of Eq.~\eqref{eq:diffnoise}] must overcome the line tension to inflate the droplet to the critical  size $r_d^*$, after which it may grow deterministically.  This is the rate-limiting step in the nucleation and growth process for an organism with selective advantage $\delta$, and  we conjecture that the probability of this step for this nonequilibrium dynamical system, i.e. the survival probability, is given by an Arrhenius  factor:
\[
P^s_{\infty} = \Omega e^{- \beta E[r_d^*]}=\Omega e^{-\frac{8 \pi  D_s  \sigma}{3D_g \delta } },\label{eq:arrhenius}
\]
where $E[r_d^*]=4 \pi  \sigma/(3 \delta) $ is the energy of the critical droplet,  $\beta$ is the inverse ``temperature'' in Eq.~\eqref{eq:effT}, and $\Omega$ is a more subtle prefactor which we now describe.

 The prefactor $\Omega$ will have two contributions. First, note that there is nothing in Eq.~\eqref{eq:arrhenius} that specifies an initial condition. In the usual (homogeneous) nucleation and growth dynamics, thermal fluctuations are responsible for initially nucleating the stable state, so the initial condition could be just the uniform metastable (blue) phase. On the other hand, for our evolutionary dynamics with the nonlinear noise that vanishes away from interfaces, all nucleating yellow cell clusters must come from yellow cells already present in the initial condition. Hence, the specific initial condition (a single yellow cell, say), will influence the prefactor $\Omega$. Secondly, it is known  that shape fluctuations around the critical droplet shape (reminiscent of entropy effects in equilibrium systems) will contribute to $\Omega$ \cite{Langer1969,voloshin}. In the thin-wall limit, these fluctuations are undulations of the interface position as shown schematically in Fig.~\ref{fig:droplet}(b).

 \begin{figure}[htp]
\centering
\includegraphics[width=0.48\textwidth]{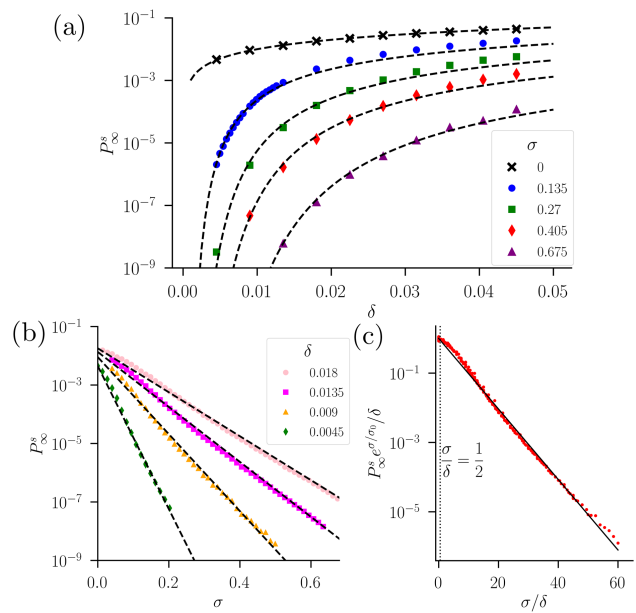}
\caption{\label{fig:survival} Survival probabilities $P^s_{\infty}$ of clusters starting from a single yellow cell in a sea of blue cells for varying antagonism $\sigma$ and selective advantage $\delta$, evaluated at  times long enough to ensure convergence using triangular lattices  of  $N=L^2$ total cells with periodic boundary conditions. (We checked that nearly identical results are obtained for $L=512,1024$.) (a) Semi-log plot of   $P_{\infty}^s$ as a function of $\delta$ for the indicated values of $\sigma$. The dashed line indicates the nucleation theory result in Eq.~\eqref{eq:nucleation}. (b)  Now $\sigma$ is varied for the indicated $\delta$. Note that in this case, the survival probability decays exponentially as a function of $\sigma$ for fixed $\delta$, as predicted by Eq.~\eqref{eq:nucleation}. (c) A data collapse of all of the survival probability data, over a wide range of $\delta$ and $\sigma$, using Eq.~\eqref{eq:survp}, where we plot $P_{\infty}^s e^{\sigma/\sigma_0}/\delta$ vs. $\sigma/\delta$ with $\tau_g\sigma_0 \equiv3D_g \tau_g\delta_0/(8 \pi D_s )\approx 0.23 $. The solid line is constructed with Eq.~\eqref{eq:survp} with best-fit parameters $D_s/D_g =0.0280\pm0.0002 $ and $\tau_g\delta_0 =0.054\pm 0.002$. The vertical dashed line shows the limit of metastability; to the left of this line, the nucleation barrier vanishes.}
\end{figure}

 In the thin-wall limit $r_d \gg \sqrt{D_s/\sigma}$,  the critical droplet can be thought of as a closed loop with a  spectrum of fluctuations which can be derived from the ``free energy'' in Eq.~\eqref{eq:freeenergy}. The fluctuations of this loop are driven by thermal noise in the case of conventional nucleation and growth. For our evolutionary dynamics, the interface shape fluctuates due to the nonlinear noise in Eq.~\eqref{eq:noise}. However, because the dynamics at the interface dominates in this thin-wall limit, we expect that the usual nucleation and growth analysis will work well, provided we evaluate the noise at the middle of the interface where $\phi=0$ (or $f=1/2$). Another possibility, similar to the approach for deriving Eq.~\eqref{eq:dropletd} from Eq.~\eqref{eq:diffpot}, is to average the nonlinear noise correlations over the stationary interface profile $\phi^*$ given by Eq.~\eqref{eq:profile}. This alternative procedure does not change our results in any substantial way (compared to evaluating the noise at $\phi=0$) and amounts to a simple rescaling of the genetic diffusion constant: $D_g \rightarrow 4D_g/5$. As we use $D_g$ as a fitting parameter when we check our analysis with simulations, the two approaches give identical results.

 Once we approximate the two-dimensional droplet of genetic material with a selective advantage $\delta$ as a fluctuating loop, we can exploit the results of  Voloshin, \cite{voloshin} who showed that in the thermal system (i.e., an Ising model below $T_c$), the nucleation rate  (including  the prefactor and the Arrhenius term) in two dimensions is completely determined by the temperature,  and the fluctuation-renormalized  condensation energy $c$ and line tension $\gamma$. The result for the nucleation rate $\Gamma$, which is necessarily proportional to the survival probability $P_{\infty}^s$ in our problem, is 
\begin{equation}
P^s_{\infty} \propto\Gamma=\frac{\beta c }{2 \pi} e^{- \beta \, \frac{ \pi \gamma^2}{c}}   \approx\frac{  \delta }{ 3\pi D_g} e^{-\frac{8 \pi  D_s  \sigma}{3D_g \delta }}, \label{eq:nucleation}
\end{equation}
where we have substituted in for the inverse temperature $\beta$ using Eq.~\eqref{eq:effT} and used the results for the line tension and condensation energy $\gamma = (2/3) \sqrt{\sigma/D_s}$ and $c = \delta/(3D_s)$, respectively.  As $\Gamma$ is a probability per unit area, the constant of proportionality must be an area related to the initial condition for our population genetics problem, which is in this case a single cell. We thus expect that this factor is some constant close to $a^2$.  As we shall see, this assumption allows standard formulas from nucleation theory to be matched smoothly onto the Kimura results for fixation probabilities when $\sigma = \gamma = 0$, and leads to an understanding of fixation probabilities in a much larger domain.

Although for a selective advantage such that $\delta > 2 \sigma$ we no longer have a potential barrier and a critical droplet size [recall that $|\delta|=2 \sigma$ defines the limit of metastability for the potential energy displayed in Fig.~\ref{fig:droplet}(a)], we can still  use the $\sigma=0$ limit to set the constant of proportionality between $\Gamma$ and $P_{\infty}^s$. In addition, we expect corrections to the nucleation formula in Eq.~\eqref{eq:nucleation}  in the exponential term, correcting the effective free energy associated with the critical droplet (to account for deviations from the thin wall limit, possible pinching off of the interface, deviations from the $\delta \ll \sigma$ limit, etc.) \cite{voloshin}. With these considerations in mind, we conjecture the following formula for the survival probability of a single yellow cell in a two-dimensional population of blue cells:
\begin{equation}
P_{\infty}^s= \frac{A_0\delta  }{D_g} \exp \left[- \frac{8 \pi  D_s  \sigma(1+ \delta/\delta_0)}{3D_g \delta } \right], \label{eq:survp}
\end{equation}
where $ \delta/\delta_0$ is a correction term that takes into account behavior away from the $\delta \ll \sigma$ limit. Note the important dependence on the spatial diffusion constant $D_s$ when $\sigma>0$, in contrast to the Kimura result for a single cell in Eq.~\eqref{eq:kimura}, which is independent of $D_s$.

We   now test this result via computer simulations. The $ D_g/A_0$ part of the prefactor  can be set by fitting to $\sigma=0$ data; we find $D_g/A_0=1.00 \pm 0.01$ in units of inverse generation time $\tau_g^{-1}$.  This result is plausible since both $\tau_g D_g$ and $A_0$ are expected to be of order $a^2$. We then determine just two dimensionless fitting parameters: $D_s/D_g =0.0280\pm0.0002  $ and $\tau_g\delta_0 =0.054\pm 0.002$, which can be found by fitting $P_{\infty}^s$ as a function of \textit{both} $\sigma$ and $\delta$ for the data with $\sigma/\delta>1/2$. The fit indicates that the spatial diffusion coefficient $D_s$ is small compared to the constant $D_g$ controlling genetic drift, consistent with the small effective population size embodied in our  simulation approach. The cells we model on a triangular simulation lattice only diffuse via cell division displacements and do not have an independent motility. With just two fitting parameters, we find good agreement between theory and simulation for $P_{\infty}^s$  as a function of both $\sigma$ and $\delta$, as shown in Fig.~\ref{fig:survival}. Note that apart from the $\delta/\delta_0$ correction, we get excellent data collapse of $P_{\infty}^s e^{\sigma/\sigma_0}/\delta$ [where $\sigma_0 \equiv 3D_g \delta_0/(8 \pi  D_s ) $] versus $\sigma/\delta$, as shown in Fig.~\ref{fig:survival}(c), where we also indicate the best fit.

Our fit to nucleation theory, shown in Fig.~\ref{fig:survival}(b,c),  reveals that the survival probability, apart from some corrections to scaling at large $\delta$,  decays exponentially with the ratio $\sigma/\delta$. This prediction is verified in the simulations. To get this good agreement, it was necessary to run the simulations until either the yellow cell cluster died out completely or spread through a large enough fraction of the population so that ultimate fixation is inevitable; this fraction was typically taken to be a quarter or a half of our simulation area.  These simulation times can be quite long because the survival probability $P^s(t)$ typically decays as a power-law in time\ $t$. This slow decay can be understood  for small $\sigma$ and $\delta$ from the behavior at the voter model point $\sigma=\delta=0$. Here, standard results \cite{Bramson} show that the survival probability in two spatial dimensions for a yellow cell cluster spawned from a single initial yellow cell decays according to $P^s(t) \sim \ln t/(\pi t)$, consistent with our simulation results from early to intermediate times before $P^s(t)$ saturates to its limiting value of $\lim_{t\rightarrow \infty} P^s(t)=P_{\infty}^s$ at long times. Some examples of the behavior of $P^s(t)$ as a function of time $t$ are shown in Fig.~\ref{fig:Inflation}(c).

An interesting aspect of the evolutionary dynamics studied here is the \textit{fixation time}, i.e., the time necessary for the yellow cell cluster to definitively sweep the population.  A rough estimate of the fixation time $t_f$ for a surviving cluster is $t_f = t_{\mathrm{nucl}}+(L-r_d^*)/v_d$, where $L$ is the linear extent of the system, $v_d = (\delta/2) \sqrt{D_s/\sigma}$ is the yellow droplet growth velocity (after it becomes larger than the critical size), and $t_{\mathrm{nucl}}$ is the ``nucleation time'' it takes for the yellow cell cluster to grow from a single cell to the critical size.  The quantity $t_{\mathrm{nucl}}$ is challenging to estimate because, unlike the late-time  droplet growth governed by the deterministic velocity $v_d$,  the early-time yellow cluster dynamics are strongly influenced by the nonlinear noise $\eta$ in Eq.~\eqref{eq:diffpot} that is ultimately responsible for overcoming the line tension induced by the antagonism $\sigma>0$.  A proper analysis would examine the evolution of the initial condition to the critical droplet shape under the stochastic evolution of Eq.~\eqref{eq:diffpot}. Here we will only  make the approximate estimate given below.

 The motion of the yellow-blue interface in the presence of noise requires a delicate analysis, and various regimes are possible depending on the magnitude of the genetic drift $D_g$ \cite{Panja}. We can make a crude estimate by assuming that a surviving subcritical yellow cluster propagates into the blue territory as a very noisy Fisher wave, driven by the genetic drift, spatial diffusion, and selective advantage $\delta>0$. The  velocity of such a wave scales according to  $v_{\mathrm{F}} \sim D_s \sqrt{\delta/D_g}$  \cite{noisyFisher}, which yields an estimated  $t_{\mathrm{nucl}} \sim r_d^*/v_{\mathrm{F}} \sim \sqrt{D_g \sigma/(D_s \delta)}\,/\delta$. This approximation yields, for example, a reasonable  $t_{\mathrm{nucl}} \approx 500$ generation times for the evolution in Fig.~\ref{fig:nucleation}(a). However, the antagonism $\sigma$ will surely modify the growth speed of the droplet and a proper analysis of the early-time dynamics would take into account the metastability for $\sigma>\delta/2$. The interplay of the antagonism and the genetic drift noise is  subtle; understanding the detailed interface motion is an active area of current research, even in one spatial dimension \cite{Birzu}.  A more detailed analysis of the fixation time is  outside of the scope of this paper. We will now analyze new features for survival probabilities on inflating spherical frontiers.

\section{Survival at Inflating Frontiers \label{sec:inflation}} 

  Spatial evolutionary dynamics can also play out on the surface of  growing tissue or at the edge of a three-dimensional microbial colony. One such example is the surface of an avascular tumor \cite{MOLSphere}. Tumors may contain  a heterogeneous distribution of strains which prefer different microenvironments and have a complicated  ecology  \cite{tumor1,tumor2,korolev1}. It is thus important to consider evolutionary interactions between cancer strains that go beyond a simple selective advantage. One  possibility is to include a frequency-dependent selection, as we do for our growth rates in Eq.~\eqref{eq:growthrates}. Such interactions are increasingly thought to be relevant for cancer strains  \cite{korolev1}. For example,  evolutionary game theory approaches suggest that a frequency-dependent selection may describe the effect of the exchange of diffusible goods, relevant for cancer ecology \cite{cancergametheory}. Thus, it is interesting to consider our antagonistic interactions in this context, as they represent a regime of frequency-dependent selection (the third quadrant in Fig.~\ref{fig:intro}).

With this motivation, consider  a spherical clump of cells (e.g., a solid tumor) with a radius $R(t)$ that grows linearly in time:
\begin{equation}
R(t) = R_0(1+t/t^*), \label{eq:radius}
\end{equation} 
where $R_0$ is the initial radius and $t^*=R_0/v$ is a time scale associated with the speed $v$ of the growing tumor frontier.   This simple microspheroid model captures many of the essential features of real tumors, especially in the early avascular stages of tumor development \cite{spheroids}. We will assume for simplicity that the frontier of the spherical tumor remains \textit{smooth}, and neglect the interesting enhancement of genetic boundary wandering associated with rough interfaces \cite{DRNPNAS,MOLMut}. In simulations, this constraint can be implemented by requiring the cell  closest to the center of the spherical cluster to divide first. This condition mimics an effective ``surface tension'' for the mass of cells, a plausible assumption since cell masses of microbes and cell tissues are known to behave as (viscoelastic) fluids under many circumstances \cite{yeastfluid,tissuefluid}. We assume that nutrients are available primarily to cells at the frontier of the growing three-dimensional tumor, so that only these cells can undergo cell divisions.

 \begin{figure}[htp]
\centering
\includegraphics[width=0.45\textwidth]{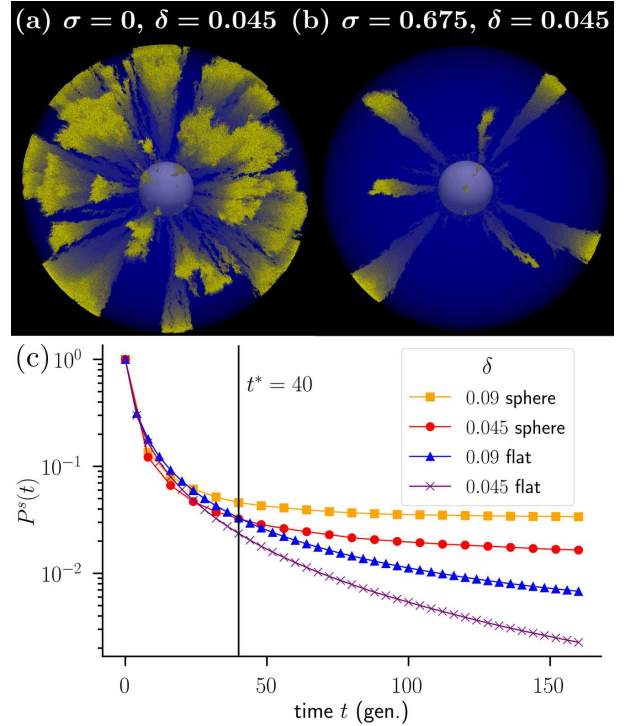}
\caption{\label{fig:Inflation} Spherical range expansions evolved from an initial radius $R_0=40a$ to a final radius $R_f=200a$ (160 generations), with $a$ the cell size, with an initial fraction of $f_0=0.02$ yellow cells for the indicated values of the antagonism $\sigma$ and yellow selective advantage $\delta$. In (a), we see that yellow sectors form and  spread (as logarithmic cones on average). The surrounding blue cells are shown in a transparent blue color. In (b), the antagonism prevents most yellow clusters from forming, despite their selective advantage $\delta=0.045$. (c) Survival probability $P^s(t)$ of a yellow cluster starting from a single yellow cell mutant at the frontier (surrounded by all blue cells) as a function of time $t$ in generations in a regular (flat) geometry (bottom two curves) and on the surface of an inflating spherical population with initial radius $R_0=40a$. Note that after a crossover time $t^*=R_0/v=40 \tau_g$, the flat and sphere cases start to  diverge. The inflating spherical geometry enhances the survival probability.  The lines connect the simulation points.      }
\end{figure}

We again consider a single mutant yellow cell appearing at the frontier of a blue spherical cluster with an initial radius $R_0$. The yellow variant grows with some selective advantage $\delta$ over the surrounding (blue) tumor cells, as would be the case for a ``driver'' mutation at a two-dimensional tumor surface \cite{driverreview}. This mutated strain might prefer its own special microenvironment and try to poison or attack its competitors, who may, in turn, produce secretions harmful to the mutant. We thus assume the curved two-dimensional blue/yellow cell interface is associated with some antagnostic interaction with a characteristic $\sigma$. An important question from the point of view of the tumor ecology is: What is the probability that the yellow cell is able to sweep the population? As illustrated in Fig.~\ref{fig:Inflation}(a,b), this quantity will depend strongly on $\sigma$, as discussed above for the flat two-dimensional environments. However, in this curved setting with growth, inflation at the tumor frontier  enhances the survival probability $P^s$, as  investigated previously with theory and simulations for the $\sigma=0$ case as a function of $\delta$ \cite{MOLSphere}.

 When $\sigma>0$, we again expect a nucleation and growth scenario, with a line tension on the curved two-dimensional tumor surface playing a key role. However, there are two important modifications to the theory in a flat environment: First, we have the inflationary effect due to the growing spherical frontier, with $R(t)$ increasing linearly with time according to Eq.~\eqref{eq:radius}.\ This factor will be especially important at late times $t \gg t^*$. Second, the curvature of the frontier will also play a role, especially if the two-dimensional critical radius $r_d^*$ of a curved yellow domain at the frontier is comparable to $R_0$ \cite{curvednuc1,curvednuc2}.  A more detailed discussion of these factors is given in Appendix~\ref{AAnalysis}.

 \begin{figure}[htp]
\centering
\includegraphics[width=0.45\textwidth]{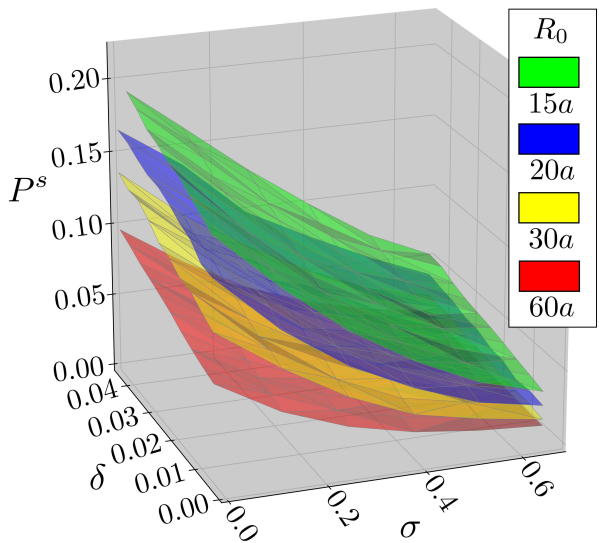}
\caption{\label{fig:infsurvp}  Simulated survival probability of single yellow cell of diameter $a$ in an initially all blue spherical cluster of cells with radius $R_0$. The survival probability $P^s$ is evaluated at a final radius of $R_f \approx 200 a$ and averaged over at least 32000 runs. The surface is generated by tessellating the simulated data points $(\sigma,\delta,P_{\infty}^s)$.    }
\end{figure}

The simulation results for the survival probability $P_s$ for a single yellow cell on an initially all blue spherical cluster of cells (after about 200 generations) is shown in Fig.~\ref{fig:infsurvp}. Note how smaller initial radii $R_0$ yield larger survival probabilities, as expected due to the increased inflationary effect.
We also see a  decay of the probability with increasing $\sigma$ and decreasing $\delta$   in Fig.~\ref{fig:infsurvp}. It is possible to derive a theoretical prediction that has similar trends. This is done in detail in Appendix~\ref{AAnalysis}. A\ key feature of the calculation is that the spatial diffusion coefficient plays a significant role, unlike the analysis for a simple selective advantage \cite{MOLSphere}. We find that for $t^* \delta \ll 1$, the long-time survival probability $P_{\infty}^s\equiv P_{\infty}^s(\sigma,\delta,t^*)$ is given by
\begin{equation}
P^s_{\infty} \approx\frac{  2A_0       }{ t^*  D_g   }        \exp\left[-\frac{4\pi t^*\left(2 \sqrt{D_s \sigma }-t^* \delta\right)}{3D_g}\right],  \label{eq:survpinfmain}
\end{equation}
where $A_0 \approx a^2$ is the initial area occupied by the yellow cell. A detailed comparison of our theoretical prediction with simulation results, along with a more detailed formula valid for larger values of $t^* \delta$ is given in Appendix~\ref{AAnalysis}.  

As discussed in more detail in the Appendix, our theoretical prediction captures the essential features of the simulated survival probability, even at finite times. We see the strong enhancement due to both the inflationary effect and the curved spherical surface.  To find better approximations, it would be necessary to study in more detail the interplay between the evolutionary dynamics and the inflation. It would also be helpful to consider the time-dependence of the survival probability. This would be an interesting topic for future research.

\section{Conclusion \label{sec:conclusion}}

  We have explored how antagonism plays a crucial role in  the spatial evolutionary dynamics of a population, leading to effects unanticipated in a simple theory of natural selection. As originally appreciated by Barton and collaborators \cite{toads, gennucleation1}, the primary effect of antagonistic interactions is to introduce a dynamical line tension between interfaces of antagonistic species. This line tension  modifies the genetic sectoring phenomenon, creating more compact sectors with more well-defined boundaries. In contrast, genetic sectors are much more diffuse when the strains are neutral, as at the voter model point $\alpha=\beta=0$ in Fig.~\ref{fig:intro}.  The transition between these two different coarsening modalities may be easily observed by tracking either the density of interfaces between species [Fig.~\ref{fig:coarsening}(b)], or by looking at the large-$k$ behavior of the structure factor associated with the identities of the species. 

Antagonism modifies the tail of the structure factor from $S(k) \sim 1/k^2$ to $S(k) \sim 1/k^3$ [Fig.~\ref{fig:SK}(a) compared to (b)].  The structure factor is experimentally accessible, as demonstrated recently in microbial colony experiments  \cite{yunker1}. However, the  $1/k^3$ behavior may be challenging to observe, as any sources of noise will obscure this tail and the structure factor will exhibit an equilibrium $1/k^2$ behavior, instead \cite{Sktail}.  Thus far, only a $1/k^2$ tail has been observed in microbial colony experiments with antagonism \cite{yunker1}.  In this previous work of McNally et al.,  the coarsening of genetic domains is compared to the Ising model below the critical temperature. We have shown that such a connection is incomplete due to the nonlinear noise that acts only at domain interfaces in the population genetics problem. The noise becomes especially important as the antagonistic strength vanishes ($\sigma \rightarrow 0$) near the voter model point $\delta = \sigma = 0$.  Moreover, varying $\sigma$ significantly alters the time-dependence of the genetic domain interface density.   Hence, it would be interesting to study, for example, the structure factor tail and the interface density  in an experiment which systematically varies the antagonistic interaction strength $\sigma$. We note that, independent of the degree of antagonism, the typical sector size of the strains should grow as $\sqrt{t}$, as is indeed observed in the microbial experiments.

We have also examined the survival probability of single mutant (yellow) cells in the presence of antagonism. Antagonism strongly suppresses the survival probability with an exponential factor which can be interpreted as an Arrhenius factor $e^{-\beta E^*}$ associated with the  formation of a ``critical cluster'' of the yellow strain, where $E^*$ is a kind of free energy of the critical cluster and $\beta$ is an effective genetic ``temperature'' given by Eq.~\eqref{eq:effT}.  An important aspect of this suppression is that the factor $\beta E^*$ is proportional to the spatial diffusion coefficient $D_s$, characterizing the  motility of the strain.  This means that, unlike the survival probability of a mutant with a simple selective advantage and no antagonism [Eq.~\eqref{eq:kimura}] which is independent of $D_s$, antagonistic interactions introduce a very strong dependence on motility. Thus, we expect that introducing a higher motility to the mutant strain will significantly suppress its survival probability. Consequently, it would be interesting to explore whether strains with antagonistic interactions evolve ways such that genetically similar offspring  ``stick together'' and thus be better able to survive.  
 
 Finally, we considered the survival probability of the mutants on the surface of inflating spherical clusters of cells, as one might find in avascular tumor growth. We showed that both the curvature of the spherical cluster and the inflationary effect due to the growth \textit{enhance} the survival probability of the mutant, as seen in Fig.~\ref{fig:infsurvp}.  We also developed a  theory of this enhancement, which focussed on effects due to the Gaussian curvature of the sphere for small times and inflationary effects for large times. In the future, it would be interesting to build a dynamical theory which would treat the full time evolution of the mutant organisms. One possibility would be to use the field-theoretic methods employed in Ref.~\cite{MOLSphere} for strains with no antagonism. We should emphasize that our statistical analysis of number fluctuations for these survival probabilities,  in both flat and curved environments, required replacing nonlinear noise correlations such as Eq.~\eqref{eq:noise} by their values \textit{at the interface} by setting $f=1/2$ [or $\phi=0$ in Eq.~\eqref{eq:diffnoise}]. Although this approximation produced reasonable agreement with our numerical simulations and is mathematically well-motivated in the thin-wall limit, future work should investigate its validity. In addition, some spatial populations have \textit{rough}, undulating frontiers which can modify these evolutionary dynamics \cite{DRNPNAS,MOLMut}. We expect frontier roughness to further enhance the survival probability of mutants, because a mutant at the frontier may be able to ``grow around'' any nearby antagonistic strains by creating a protrusion into the surrounding empty space.  

\qquad

\begin{acknowledgments}
M.O.L. thanks B. T. Weinstein and E. M. Horsley for helpful discussions. D.R.N. benefited from discussions with R. Benzi. Computational support was  provided by the University of Tennessee and Oak Ridge National Laboratory's Joint Institute for Computational Sciences. M.O.L. gratefully acknowledges partial funding from the Neutron Sciences Directorate (Oak Ridge National Laboratory), sponsored by the U.S. Department of Energy, Office of Basic Energy Sciences.
Work by D. R. N. was supported by the National Science Foundation, primarily through Grant No. DMR-1608501, and through the Harvard Materials Science and Engineering Center, through Grant No. DMR-1420570.
\end{acknowledgments}

\appendix

\section{\label{AAnalysis}Calculating the Survival Probability with Inflation}

In this Appendix we give additional details related to calculating the long-time survival probability $P_s^{\infty}(\sigma,\delta)$ of a single mutant yellow cell with selective advantage $\delta$ evolving in a sea of blue cells at the surface of a growing spherical cluster of cells.  A convenient coordinate system for describing these dynamics is one in which each point $\mathbf{y}$ on the population frontier at time $t$ is traced back to a reference position $\mathbf{x}=R_0 \mathbf{y}/R(t)$ on the initial frontier. In this coordinate system, the dynamics evolves according to Eq.~\eqref{eq:steppingstone} but with the altered spatial and genetic diffusion constants $D_{s,g} \rightarrow D_{s,g}/(1+t/t^*)^2$. A detailed analysis for $\sigma=0$, the case of zero antagonism,  gives us an inflationary analog of the Kimura formula (see \cite{MOLSphere} for details) for the survival probability of a single yellow cell on an initially all blue cell spherical frontier:
\begin{align}
P^s_{\infty}(\sigma=0,\delta,t^*) & = 1- \exp\left[- \frac{  4A_0e^{-t^*\delta /2}}{D_g \delta (t^*)^2G( t^*\delta/2)} \right] \nonumber \\
& \approx\left[\frac{   e^{-t^*\delta /2}}{  ( t^*\delta/2)^2G( t^*\delta/2)} \right]\frac{  \delta A_0 }{D_g   } , \label{eq:kimurainf}
\end{align}
with $G(x)=\int_x^{\infty}z^{-2}e^{-z}\mathrm{d}z$ an incomplete gamma  function. Note that Eq.~\eqref{eq:kimurainf} reduces to Eq.~\eqref{eq:kimura} in the $ t^*\delta \rightarrow \infty$ limit of a planar frontier. It can be checked  that the inflating survival probability at long times is always larger than the non-inflating one: $P^s_{\infty}(0,\delta,t^*)> P^s_{\infty}(0,\delta,\infty)$  for any $t^*<\infty$. As expected, the inflating environment on the surface of the sphere always enhances survival. Moreover, note that the survival probability no longer vanishes as the selective advantage $\delta \rightarrow 0$. Indeed, inflation can rescue neutral mutations from extinction, as discussed in more detail in \cite{MOLSphere}.

  Note that the formula in Eq.~\eqref{eq:kimurainf} does not include the spatial diffusion coefficient $D_s$, much like the original Kimura formula in Eq.~\eqref{eq:kimura}. The effects of inflation (within the approximations discussed in Ref.~\cite{MOLSphere}) are entirely captured by the increase in local population size and subsequent decrease in the genetic drift coefficient $D_g$. When we include antagonistic interactions, however, we know that the analysis will have to change substantially as the dynamical line tension $\gamma$ introduces an explicit dependence on $D_s$.  So, to proceed in the case $\sigma>0$, we now consider the nucleation and growth scenario on the surface of an inflating sphere.

We first return to  the   deterministic dynamics of a cluster of yellow cells.  We expect that the overall spherical growth will deterministically inflate the initial droplet area by a factor of $(1+t/t^*)^2$, while inflating
its perimeter by a factor of $1+t/t^*$.  The deterministic dynamics in Eq.~\eqref{eq:dropletd} then changes and we find that $r_d(t)$ has a \textit{modified critical radius} $r_d^*(t^*)$, where $r_d(t) \rightarrow \infty$  if $r_d(t=0)>r_d^*(t^*)$ (and $r_d(t) \rightarrow 0$, otherwise), which reads
\begin{equation}
r_d^*(t^*)=\left[ \frac{2\pi t^*\delta   }{9} e^{\frac{2 \pi  t^*\delta }{9}} E_{1/3}\left(\frac{2 \pi  t^*\delta  }{9}\right)\right]\sqrt{\frac{4D_s\sigma}{\delta^2}}, \label{eq:rcritinf}
\end{equation}
where $E_{1/3}(x)=\int_1^{\infty}\mathrm{d}z\,e^{-zx}/z^{1/3}$ is an exponential integral function. The expression correctly reduces to the non-inflating result when the curvature effects vanish in the limit $t^* \rightarrow  \infty$: $r_d^*(t^*\rightarrow \infty)=\sqrt{4D_s \sigma/\delta^2}$. Moreover, we have $r_d^*(t^*)<r_d^*(t^* \rightarrow \infty)$ for any $t^*$, so that inflation always makes it easier for the yellow cluster to achieve a critical size  (inflation protects against genetic drift).

  \begin{figure}[htp]
\centering
\includegraphics[width=0.45\textwidth]{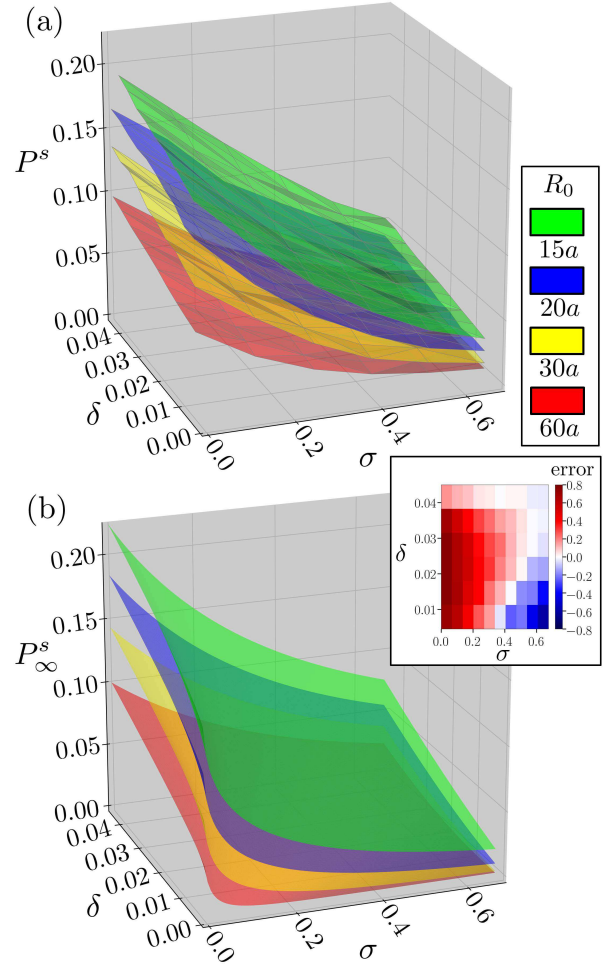}
\caption{\label{fig:infsurvp2} (a) A reproduction of the simulated survival probability of single yellow cell of diameter $a$ show in Fig.~\ref{fig:infsurvp}.  (b) Theoretical predictions for the long-time survival probability $P_{\infty}^s$ [see Eq.~\eqref{eq:survpinf}] for the same initial radii $R_0$, with fit parameters $D_g=33\pm4$ and $D_s=0.43\pm0.06$ (in units where $a=\tau_g=1$). We also fit the parameter $\tau_gD_g/A_0=0.983\pm0.008$ with the $\sigma=0$ data. The theoretical prediction matches to the simulations reasonably well, except near the two boundaries $\delta=0$ and $\sigma=0$ of the plot, where either the critical radius is larger than the sphere radius ($\delta=0$ boundary) or where we no longer have metastability ($\sigma=0$ boundary). This breakdown is illustrated in the inset, which  shows the relative error $(P^s-P^s_{\infty})/P^s$ between simulation and theory for $R_0=30 a$. }
\end{figure}

So far we have only been concerned with the effects of inflation at the frontier. However, the spherical frontier also has a nonzero Gaussian curvature. Consequently, the flat space formulas for the perimeter and area of a circle must be modified. For a fixed perimeter  of a nucleating yellow cluster, the area enclosed will be \textit{larger} on a spherical surface than on a flat surface due to  the positive curvature. On a spherical surface of radius $R_0$, this geometric effect modifies the energy $E\equiv E[r_d]$ [see Eq.~\eqref{eq:denergyflat}]  and the critical radius $r_d^*$ \cite{curvednuc1} as follows:
\begin{equation}
\begin{cases}\bar{r}_d^*=R_0 \arctan\left[\dfrac{r_d^*}{R_0}\right]<r_d^* \\[10pt]
\bar{E}=2 \pi R_0 \left[\gamma \sin\left[\dfrac{\bar{r}_d}{R_0}\right]-2cR_0\sin^2\left[\dfrac{\bar{r}_d}{2R_0}\right] \right]
\end{cases}, \label{eq:curved}
\end{equation} 
where the bars indicate the quantities on the sphere. Here, the droplet radius $\bar{r}_d$ is measured \textit{along} a great circle of the sphere surface from the droplet center. Note that the critical radius $\bar{r}^*_d$ on the sphere is always smaller than the flat surface value $r_d^*$.
Hence, \textit{both} the inflationary expansion \textit{and} the curved droplet at the frontier will \textit{enhance} the yellow sector survival probability. We see from the simulation results  in Fig.~\ref{fig:infsurvp2}(a) that the survival probability is in fact significantly enhanced for smaller $R_0$ for all values of $\sigma$ and $\delta$.

 If the fate of the yellow sector is decided (by a successful nucleation event or extinction) early compared to the inflationary time scale $t^*$, then we may use the non-inflating droplet energy $\bar{E}$ on a spherical surface given in Eq.~\eqref{eq:curved}    to calculate the survival probability, since we expect $R(t) \approx R_0$ at early times. This approximation is reasonable provided the genetic drift is sufficiently strong, which we expect to be the case for a thin layer of growing cells at the population frontier, as in our simulations. The idea here is that surviving yellow cell  clusters will reach size $\bar{r}_d^*$ early when $R(t) \approx R_0$ and then deterministically spread through the population as both inflation and the selective advantage $\delta$ take over. Thus,  evaluating $\bar{E}$ at $\bar{r}_d=\bar{r}_d^*$ and matching to the (inflating) Kimura formula in Eq.~\eqref{eq:kimurainf} at $\sigma=0$, we find that the inflationary survival probability with antagonism is approximately
\begin{align}
&P^s_{\infty}(\sigma,\delta,t^*) \approx\,\frac{  A_0      }{D_g   }  [ p_1( t^*\delta/2)]\delta\times    \, \nonumber \\ & \exp\left[{-\frac{ 4 \pi R_0^2 \delta  }{3D_g} \left[ \frac{\frac{4D_s  \sigma }{( R_0\delta)^2} p_2(t^*\delta )+1}{\sqrt{1+\frac{4D_s\sigma}{( R_0  \delta)^2}\left[p_2( t^*\delta) \right]^2}}-1\right]}\right], \label{eq:survpinf}
\end{align}
where we found it convenient to define two new functions $p_1(x)\equiv e^{-x}[x^2G(x)]^{-1}$ and $p_2(x)\equiv 2\pi x e^{2 \pi x/9} E_{1/3}(2\pi x/9)/9$ in terms of the incomplete gamma function $G(x)$ and the exponential integral function $E_{1/3}(x)$.
Note that the effects of the curved front are negligible when $ \sigma D_s  \ll (R_0\delta )^2$. Our simulations are in the regime where $\sigma D_s$ is comparable to $(R_0 \delta)^2$. An approximation to Eq.~\eqref{eq:survpinf} for  $\delta t^* \ll 1$ is given by Eq.~\eqref{eq:survpinfmain} in the main text.

Unlike the flat, two-dimensional population simulations, our analysis here is limited because we only had the resources to simulate our spherical clusters out to a maximum radius of 200 cell diameters. Although this is a rather large cluster (approximately 40 million cells in the Bennett model of closely packed hard spheres \cite{bennett}), the survival probability may still be decaying and we  expect in general simulated probabilities larger than the prediction in Eq.~\eqref{eq:survpinf}. Nevertheless, we can get a sense of the behavior of the survival probability by proceeding as in the flat case. We first use the $\sigma=0$ data to fit the parameter $\tau_gD_g/A_0 \approx 0.983 \pm 0.008$. Since our Bennett packing model simulation is implemented such that $v=a/\tau_g=1$, we then just need the fitting parameters $D_g=33\pm4$ and $D_s=0.43\pm0.06$ (in units where $a=\tau_g=1$) to fully specify $P_{\infty}^s$ in Eq.~\eqref{eq:survpinf}. The results are shown in Fig.~\ref{fig:infsurvp2}(b).  Our fitted genetic drift magnitude $D_g$ is large compared to the flat population, probably due to the incompatibility of comparing $P_{\infty}^s$ to a simulated survival probability at a finite time. In addition, we do generally expect different values for our parameters compared to the flat populations because the Bennett cluster algorithm generates a  disordered sphere packing.  The inset to Fig.~\ref{fig:infsurvp2} shows the relative error between the theoretical $P_{\infty}^s$ and the simulated $P^s$ for the $R_0=30 a$ case.  The theoretical $P_{\infty}^s$ is generally less than the simulated $P^s$, as we might expect from the finite time of the simulation. Note that the theory does not match the simulations as well for small $\delta$ and for small $\sigma$, as can also be seen in the inset in Fig.~\ref{fig:infsurvp2}.  This anomaly arises because we are outside the region of metastability for small $\sigma$. Also, at small  values of $\delta$,  the critical yellow cluster size becomes large and the fate of the yellow cell cluster will no longer by decided at early times $t \ll t^*$, so the approximations described above break down.

\bibliographystyle{apsrev}
\bibliography{antbib}{}
\end{document}